\documentclass[journal]{IEEEtran}
\usepackage{graphicx}
\usepackage{multirow}
\ifCLASSINFOpdf
\else
\fi
\begin{document}
%
\title{Clinical Brain-Computer Interface Challenge 2020 (CBCIC at WCCI2020): Overview, methods and results}
%
%
%

\author{Anirban~Chowdhury,~\IEEEmembership{Member,~IEEE,}
        and~Javier~Andreu-Perez,~\IEEEmembership{Senior Member,~IEEE}
\thanks{A. Chowdhury and J. Andreu-Perez are with the School of Computer Science and Electronic Engineering, University of Essex, Colchester,
Essex, CO4 3SQ United Kingdom e-mail: (a.chowdhury@essex.ac.uk; javier.andreu@essex.ac.uk).}
\thanks{Manuscript received March XX, 2021; revised XXXX XX, 20XX.}}

%
%

\markboth{IEEE Transactions on Medical Robotics and Bionics,~Vol.~XX, No.~X, March~20XX}%
{Shell \MakeLowercase{\textit{et al.}}: Review of the Clinical BCI Challenge I at WCCI 2020}
%



\maketitle

\begin{abstract}
In the field of brain-computer interface (BCI) research, the availability of high-quality open-access datasets is essential to benchmark the performance of emerging algorithms. The existing open-access datasets from past competitions mostly deal with healthy individuals' data, while the major application area of BCI is in the clinical domain. Thus the newly proposed algorithms to enhance the performance of BCI technology are very often tested against the healthy subjects' datasets only, which doesn't guarantee their success on patients' datasets which are more challenging due to the presence of more non-stationarity and altered neurodynamics. In order to partially mitigate this scarcity, Clinical BCI Challenge aimed to provide an open-access rich dataset of stroke patients recorded similar to a neurorehabilitation paradigm. Another key feature of this challenge is that unlike many competitions in the past, it was designed for algorithms in both with-in subject and cross-subject categories as a major thrust area of current BCI technology is to realize calibration-free BCI designs. In this paper, we have discussed the winning algorithms and their performances across both competition categories which may help develop advanced algorithms for reliable BCIs for real-world practical applications.
\end{abstract}


%
\IEEEpeerreviewmaketitle

\section{Introduction}
%
%
%
%
\IEEEPARstart{T}{hrough} the last two decades, the BCI technology has seen rapid advancement, which has brought it from the lab environment to the brink of practical uses. The advancement is made possible by the work of many researchers around the world who are actively working in this area. Like other areas of technology, a massive impetus in its growth is the availability of huge computing resources at an affordable price and the tremendous development in machine learning and artificial intelligence in the last 10 years. Earlier BCIs, which were dependent on the time consuming operant conditioning where the users are trained to produce a specific brain-wave modulation, have been replaced by adaptive/co-adaptive BCIs where the machine is trained to adapt to the changes in the brain-wave modulations. At the same time, the user also sometimes tries to stabilize their brain-wave pattern by focusing more on the BCI task. Nevertheless, any significant advent in the BCI technology was always accompanied by increasingly efficient algorithms that intend to produce reliable decoding of the brain signals associated with communication and control commands. 

\begin{figure}[ht]
	\centering
	\includegraphics[width=0.48\textwidth,keepaspectratio]{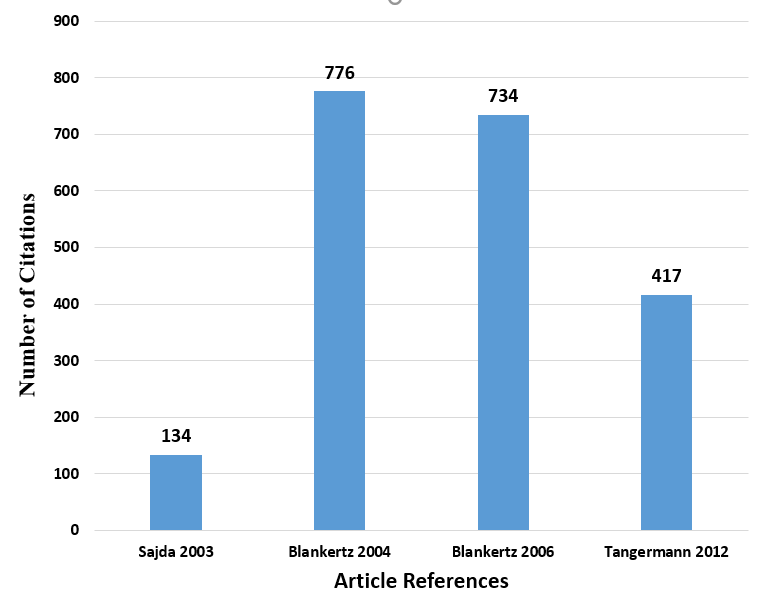}	
	\caption{The citation counts of the overview articles of BCI Competition I-IV. The article references are as follows: Sajda 2003~\cite{Sajda2003},  Blankertz 2004~\cite{Blankertz2004}, Blankertz 2006~\cite{Blankertz2006},and Tangermann 2012~\cite{Tangermann2012} refers to the overview articles of BCI Competition I, II, III, and IV, respectively.}
	\label{ref:citationCounts}
\end{figure}

A systematic development of such algorithms was made possible largely due to the availability of open access datasets generated during several BCI competitions such as BCI Competition I-IV organized between year 2003 and 2008. These competitions provided high quality neuroscientific data across different BCI paradigms (e.g. steady-state visual-evoked potentials (SSVEP), P300, sensory motor rhythms (SMR), error-related negativity responses (ERN), and movement-related cortical potentials (MRCP)) and brain-signal modalities (e.g. electroencephalogram (EEG), magnetoencephalogram (MEG), Electrocorticogram (ECoG), and functional Near Infrared Spectroscopy (fNIRS)). The relevance of such competitions can be appreciated from the number of citations the associated overview articles (\cite{Sajda2003,Blankertz2004,Blankertz2006,Tangermann2012}) have achieved over the years, which is also an indication of how widely these datasets were used by researchers around the world. The graph in Fig.~\ref{ref:citationCounts} shows the total citations acquired by the overview articles of BCI competition I-IV as of date 6$^{th}$ October 2020 according to the google scholar indexing.

The first BCI competition was organized as a part of Brain Computer Interface Workshop at the Neural Information Processing Systems in 2001. The challenge was to predict the executed and imagined button press by the left/right hand from the three EEG datasets which were presented on such tasks. Two of these datasets were recorded in open-loop while the third one was recorded in closed-loop. Altogether 10 submissions were received, out of which the winning algorithm was the recurrent
neural network for dataset 1 with a fraction correct (FC) score of 0.96. The winning entry for dataset 2 took the approach of combined feature from event-related potentials (ERP), adaptive autoregressive models (AR), and common spatial pattern (CSP), after which a Fisher discriminant was used for classification~\cite{Sajda2003}. 

In the overview paper on BCI competition II~\cite{Blankertz2004}, Blankertz and colleagues raised three fundamental issues regarding the EEG based BCI, which were in desperate need for answers. These were the issues of data quality, the generalizability of the extracted features against the overfitting nature of the classifiers, and whether the classifier performance is transferrable from the training without feedback to the online feedback stage. In this competition, there were altogether 6 datasets, namely Ia, Ib, IIa, IIb, III, and IV, contributed by four renowned laboratories around the world. The datasets Ia and Ib were based on slow cortical Potential (SCP) paradigm where the BCI task was to move a cursor up or down by self-modulation of the SCP positively/negatively. The dataset Ia dealt with the data from healthy individuals, while dataset Ib dealt with the data from the patients paralysed due to amyotrophic lateral sclerosis (ALS). Although the winning entry for the dataset Ia achieved an error rate of 11.3\%, the best result obtained for dataset Ib was nearly at chance level with an error rate of 45.6\%, which may primarily be influenced by the low-quality of data. The dataset IIa was based on the sensorimotor rhythm (SMR) paradigm consisting of 4-classes related to cursor control from 3 subjects. Out of the 5 submissions in this category, the best performance was achieved using regularized linear discriminant analysis on bandpass filtered common spatial patterns, which yielded an accuracy of 71.8\% against a chance level of 25\%. The dataset IIb was based on a P300 ERP paradigm from one subject. Interestingly 5 out of 7 submissions for this dataset achieved 100\% accuracy. Dataset III consisted of a left and right hand MI task from a healthy female subject. The challenge was to achieve the maximum mutual information with minimum delay. The best result was obtained as mutual information 0.61 bits among all the 9 submissions. Dataset IV consisted of a self-pace tapping task of four different keys and the objective was to predict the labels with maximum accuracy. The winning entry for this dataset achieved an error rate of 16\% using common spatial subspace decomposition (CSSD) with Fisher discriminant analysis (FDA)~\cite{Wang2004}.  

The third BCI Competition was more focused on advanced problems such as nonstationarity of brain signals, inter-session and inter-subject transfer learning, asynchronous BCI, and learning under limited training examples~\cite{Blankertz2006}. There were altogether 8 datasets in this competition all of which were related to MI signals except dataset II. The challenge in dataset I was to achieve robust classifier performance under nonstationarity due to changing affective states of the user and changing experimental conditions (such as slight variation of electrode position, impedance etc.). Electrocorticogram (ECoG) data from a subject who suffered from epilepsy were recorded on two different days for this dataset while the subject performed MI task using left hand finger and tongue. The winner out of a total 27 submissions, achieved 91\% accuracy using a combination of CSSD and mean waveforms calculated from the bandpower features while the feature selection was done using Fishers discriminant analysis. A linear support vector machine (SVM) was used for classification purpose. Dataset II was based on a P300 speller paradigm from 2 subjects recorded over 5 sessions. Out of a total of 10 submissions for this dataset the winner achieved an accuracy of 96.5\%. A four class MI classification task (left and right hand, foot, and tongue movement) was assigned for dataset IIIa which comprised of the data from 3 subjects. The best performance with kappa 0.79 was observed for the method comprising of multiclass CSP and SVM. The problem of continuous binary classification was given in the dataset IIIb. The best result was achieved using ERP and ERD features on a probabilistic classification model with a kappa value of 0.32. Dataset IV dealt with motor imagery signals in various conditions such as with limited training examples (IVa), asynchronous BCI (IVb), and a three class classification involving a resting state class (IVc). The winning algorithm achieved 94.2\% accuracy in dataset IVa, while for IVc the best mean square error was 0.3. The ERD features calculated by CSP/CSSD were prominent among other top performing features while LDA and FDA were popular among the classifiers. The dataset V took a mixed imagery approach where two classes were based on self-paced left and right hand movement and another class was based on word generation. Out of 26 submission for this dataset the winning entry yielded an error of 31.3\%. A key observation from BCI competition III was that due to very high inter-subject variability specialized algorithms or signal processing pipelines are required for achieving the best BCI performance. 

The BCI Competition IV was especially focused on the oscillatory brain activity associated with different motor tasks such as MI and motor execution (ME)~\cite{Tangermann2012}. This is because a major application area of BCI technology is in rehabilitating the motor impairments of neurologically disabled people by promoting neuroplasticity~\cite{Chowdhury2018IEEEJBHI,Chowdhury2020IEEEAccess}. There were 5 different datasets in this competition across three different brain modalities such as EEG, MEG, and ECoG. The challenges varied from asynchronous BCI designs, multiclass continuous classification, inter-session transfer learning, directional decoding using the same hand, and the discrimination of individual finger movements. Out of the 24 submissions for dataset 1, where the challenge was to predict the left and right hand MI along with the resting state continuously in an asynchronous BCI paradigm, the winning team used filter-bank CSP (FBCSP) based features using mutual information for feature selection. The average mean square error (MSE) achieved was 0.382~\cite{Zhang2012}. The dataset IIa is one of the most famous datasets as it involves a greater challenge of multiclass classification with 4 classes in continuous mode. The best performing algorithm, in this case, was again FBCSP with a multiclass extension where Naive Bayes Parzen Window classifiers were used for the classification purpose~\cite{Ang2008}, yielding a kappa of 0.57. The analysis of multiple hand-crafted features in combination with fast learning classifiers has been a popular approach in BCI \cite{gupta2019}. The dataset IV-2b was another highly used dataset which challenged the robustness of the algorithm for session-to-session transfer of continuous binary (left vs. right hand MI) classifier. In this case, also the FBCSP algorithm gave the superior performance over the others with a kappa of 0.6~\cite{Ang2012}. The idea behind dataset III was to explore the feasibility of directional decoding using MEG signals which is important in the context of rehabilitating people with hand disability due to neurologic conditions. Although only four submissions were in this category, the winning algorithm used a large variety of features such as statistical features, time and frequency domain features, and wavelet coefficients followed by a genetic algorithm to optimize the classification accuracy from a combination of linear SVM and LDA classifiers. The best average accuracy across all the subjects was 53\% which was well over the chance level given there were four directional classes to be classified. Dataset IV dealt with ECoG signals related to the flexion of five different fingers where the correlation coefficient ($r$) between the actual and predicted classes was set as the evaluation criterion. The winning algorithm achieved an $r$ of 0.46. A summary of all the winning methodologies for various datasets based on different protocols from the past 4 BCI competitions can be found in Table~\ref{tab:prevResults}.


\begin{table*}[htb]
\caption{Results from the previous BCI competitions I-IV}
\centering
\label{tab:prevResults}
\begin{tabular}{|c|c|c|c|}
\hline
\textbf{Competition} & \textbf{Dataset} & \textbf{Protocols}                                                                                          & \textbf{Winning   Methodologies}                                                                                                   \\ \hline
\multirow{2}{*}{I}   & Dataset 1        & EEG Self-Paced Key Typing                                                                                   & \begin{tabular}[c]{@{}c@{}}Recurrent\\      Neural Network (RNN)\end{tabular}                                                      \\ \cline{2-4} 
                     & Dataset 2        & EEG Synchronized Imagined Movement                                                                          & ERP+AR+CSP+FDA                                                                                                                     \\ \hline
\multirow{5}{*}{II}  & Datasets Ia      & Self-regulation of SCPs                                                                                     & DC potential+High Beta bandpower+LDA                                                                                               \\ \cline{2-4} 
                     & Datasets Ib      & Self-regulation of SCPs                                                                                     & Wavelet transform+Stepwise LDA                                                                                                     \\ \cline{2-4} 
                     & Dataset IIa      & \begin{tabular}[c]{@{}c@{}}Seld-Regulation of Mu and/or\\      Central Beta Rhythm\end{tabular}             & Bandpass filtering+CSP+Regularized LDA                                                                                             \\ \cline{2-4} 
                     & Dataser III      & Motor Imagery                                                                                               & Mutual Information                                                                                                                 \\ \cline{2-4} 
                     & Dataset IV       & Self-paced tapping                                                                                          & CSSD+FDA                                                                                                                           \\ \hline
\multirow{6}{*}{III} & Dataset I        & ECoG related to MI                                                                                          & Bandpower+CSSD+FDA+SVM                                                                                                             \\ \cline{2-4} 
                     & Dataset IIIa     & EEG related to MI                                                                                           & Multiclass CSP+SVM                                                                                                                 \\ \cline{2-4} 
                     & Dataset IIIb     & EEG related to MI                                                                                           & \begin{tabular}[c]{@{}c@{}}ERP+ERD+Propabilistic classifier+\\ Accumulative Classifier\end{tabular}                                \\ \cline{2-4} 
                     & Dataset IVa      & EEG related to MI                                                                                           & \begin{tabular}[c]{@{}c@{}}Combination of CSP,  AR and Temporal Waves \\ of the readiness potential+LDA\end{tabular}               \\ \cline{2-4} 
                     & Dataset IVc      & MI with asynchronous protocol                                                                               & ERD by CSSD+FDA                                                                                                                    \\ \cline{2-4} 
                     & Dataset V        & \begin{tabular}[c]{@{}c@{}}Multiclass continuous EEG for \\ Motor and Cognitive Imagery\end{tabular}        & Power Spectral Density (PSD)+FDA                                                                                                   \\ \hline
\multirow{5}{*}{IV}  & 1                & \begin{tabular}[c]{@{}c@{}}EEG based 2-Class MI, \\ uncued classifier application\end{tabular}              & Filter Bank CSP (FBCSP)+Neural Network                                                                                             \\ \cline{2-4} 
                     & 2a               & \begin{tabular}[c]{@{}c@{}}EEG based 4-Class MI, continuous\\      classifier application\end{tabular}      & FBCSP+Naive Bayes Parzen Window classifier                                                                                         \\ \cline{2-4} 
                     & 2b               & \begin{tabular}[c]{@{}c@{}}EEG based MI, session-to-session\\      transfer and eye artifacts\end{tabular}  & FBCSP+Naive Bayes Parzen Window classifier                                                                                         \\ \cline{2-4} 
                     & 3                & \begin{tabular}[c]{@{}c@{}}MEG based Decoding directions of\\      finger/hand/wrist movements\end{tabular} & \begin{tabular}[c]{@{}c@{}}wavelet coefficients+Genetic Algorithm+\\ a combination of SVM and LDA\end{tabular}                     \\ \cline{2-4} 
                     & 4                & \begin{tabular}[c]{@{}c@{}}ECoG Discrimination of movements \\ of individual finders\end{tabular}           & \begin{tabular}[c]{@{}c@{}}Band pass filtering+pair-wise feature selection+\\ Linear regression using Wiener solution\end{tabular} \\ \hline
\end{tabular}
\end{table*}


From the past competitions, it is evident that most of the datasets are based on EEG signals acquired in an SMR paradigm, although the subjects are mostly healthy individuals. This is a major drawback in evaluating the performance of the algorithms in real-world applications as the target users of the BCI technology are mainly in the clinical domain. Another limitation in these datasets is the number of subjects, as the average number of subjects per dataset is only 3.69$\pm$2.93. This poses an important problem while comparing a newly proposed algorithm against the previous state-of-the-art as the statistical significance of the results is difficult to be tested in such a small sample size. Therefore, only a very small subset of these datasets such as BCI Competition IV 2a, 2b are mainly popular in literature where the sample size is a bit higher ($n$=9). Thus there was a need for a comprehensive clinical dataset for SMR based BCI, which is openly accessible with a sufficient number of subjects who are the actual target users of the BCI technology. For partial mitigation of this necessity, we organized the Clinical BCI Challenge (CBCIC) as part of the World Congress on Computational Intelligence (WCCI 2020) in 2020 at Glasgow, United Kingdom. The dataset provided the left and right hand motor attempt EEG data from 10 different hemiparetic stroke patients. A number of teams from all around the world participated in this competition submitting different emerging and traditional algorithms to meet the challenge. In this paper, we have described the dataset and the challenge, given an overview of the competition, and discussed the results and methodologies which are applied to achieve those results. We hope this will provide a benchmark for developing advanced data processing pipelines, which would spawn the next generation of neurorehabilitative BCI designs and simultaneously serve as a rich and challenging dataset to make a statistical comparison between newly proposed algorithms in the field of SMR BCI~\cite{Roy2020JNE,Roy2020FIN,PramodChowdhury2021TIM}.

\section{Overview of the Competition}
The competition was organized as a part of IEEE World Congress on Computational Intelligence (WCCI) 2020, held in Glasgow, United Kingdom in July 2020. It was under the section of IEEE Congress on Evolutionary Computation (IEEE CEC) competitions with the competition code CEC-C1\footnote{https://wcci2020.org/competitions/} and a corresponding website\footnote{https://sites.google.com/view/bci-comp-wcci/}. 

\subsection{The Challenge}
An EEG dataset from 10 individuals, having a clinical condition of hemiparesis due to stroke, were provided. The dataset was generated out of a neurorehabilitative BCI experimental paradigm composed of sensorimotor rhythm (SMR) associated with two classes of left and right hand grasp-attempts. The challenge was single trial decoding of the brain signals into respective classes in two different approaches: 1) within-subject decoding and 2) cross-subject decoding. In the within-subject approach, the classifier was trained using the training data of the same subject and evaluated on the test data of that subject. In the cross-subject decoding, the training data from the first 8 subjects (P01-P08) were used to train the classifier, which was evaluated on the test data of the last two subjects (P09 and P10). The objective of cross-subject decoding was to evaluate the robustness of the algorithm against the inter-subject variability. It is to be noted that the training data of the subjects P09 and P10 were not given in order to avoid any effect of the same subject influence over the cross-subject approach. All the participants are asked to submit the results for both the categories (within-subject and cross-subject) so that the question of whether a single algorithm is capable enough to get the best performance in both the approaches can be investigated. It is worth mentioning that only the training labels were provided to the participants while the test labels were kept hidden. The ranking of the submissions was done on the basis of the kappa ($\kappa$) values of the classification.

\begin{figure}[ht]
	\centering
	\includegraphics[width=0.48\textwidth,keepaspectratio]{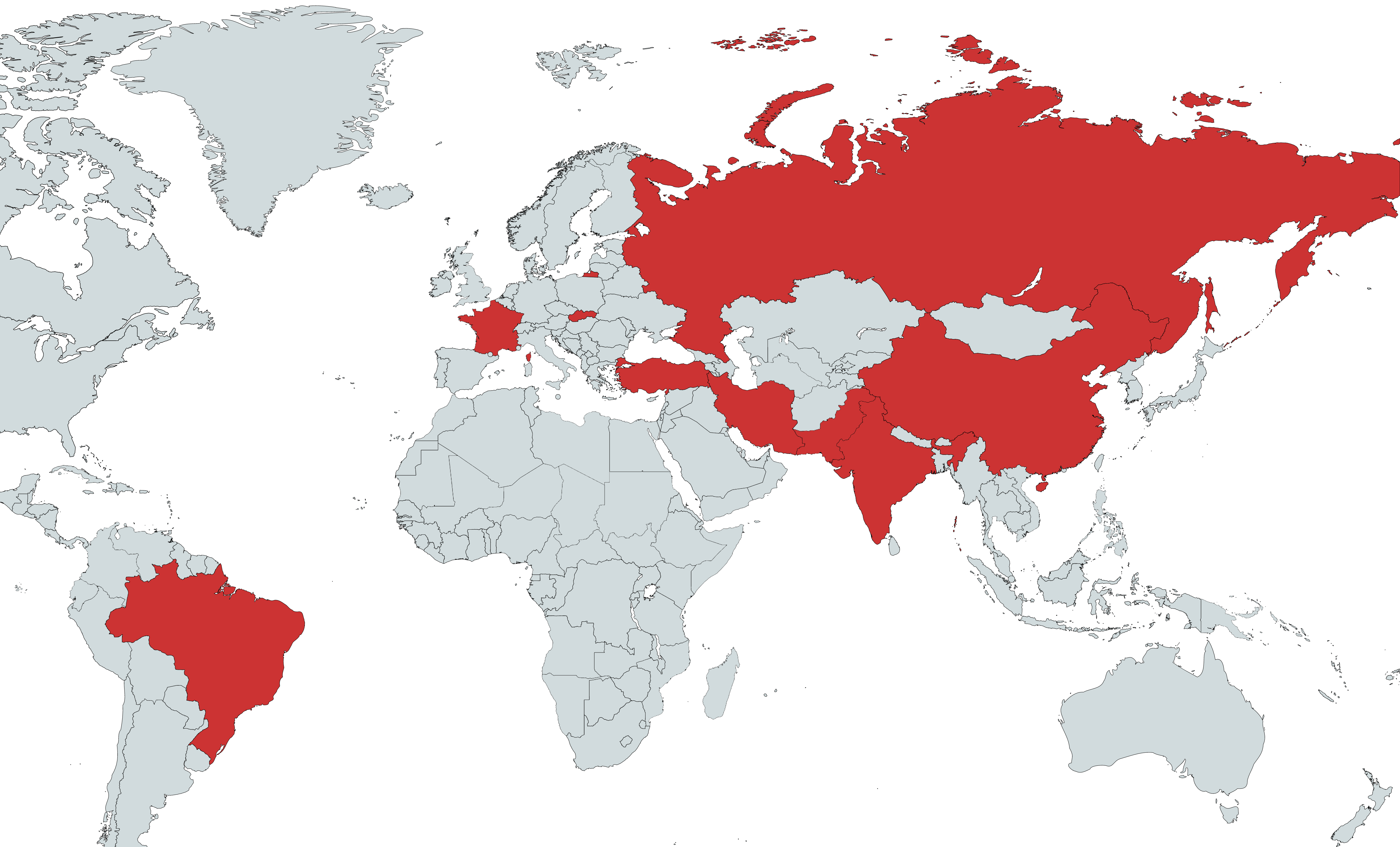}	
	\caption{Geographic distribution of the participants.}
	\label{fig:map}
\end{figure}

\begin{figure}[ht]
	\centering
	\includegraphics[width=0.48\textwidth,keepaspectratio]{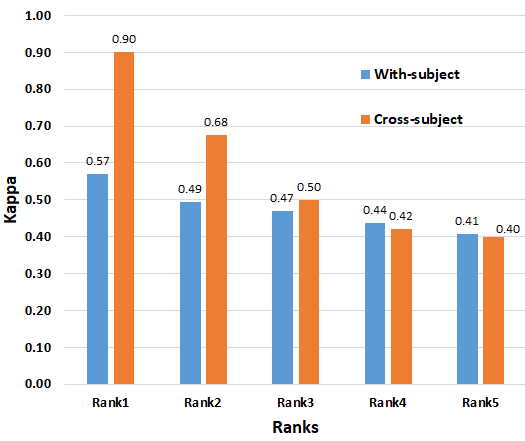}	
	\caption{The comparison of kappa values between the top 5 ranks.}
	\label{fig:interCatComp}
\end{figure}

\begin{figure}[ht]
	\centering
	\includegraphics[width=0.45\textwidth,keepaspectratio]{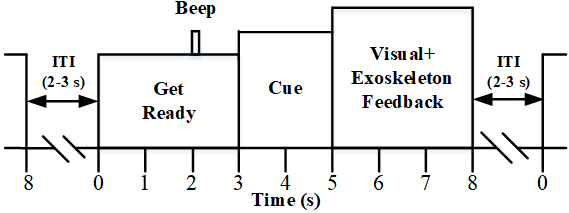}	
	\caption{The timing diagram of a single trial in the online feedback stage. In the training stage the cue lasted for the whole 5~s period after the cue as there was no feedback.}
	\label{fig:timingDiagramTrial}
\end{figure}



\begin{figure*}[ht]
	\centering
	\includegraphics[width=0.75\textwidth,keepaspectratio]{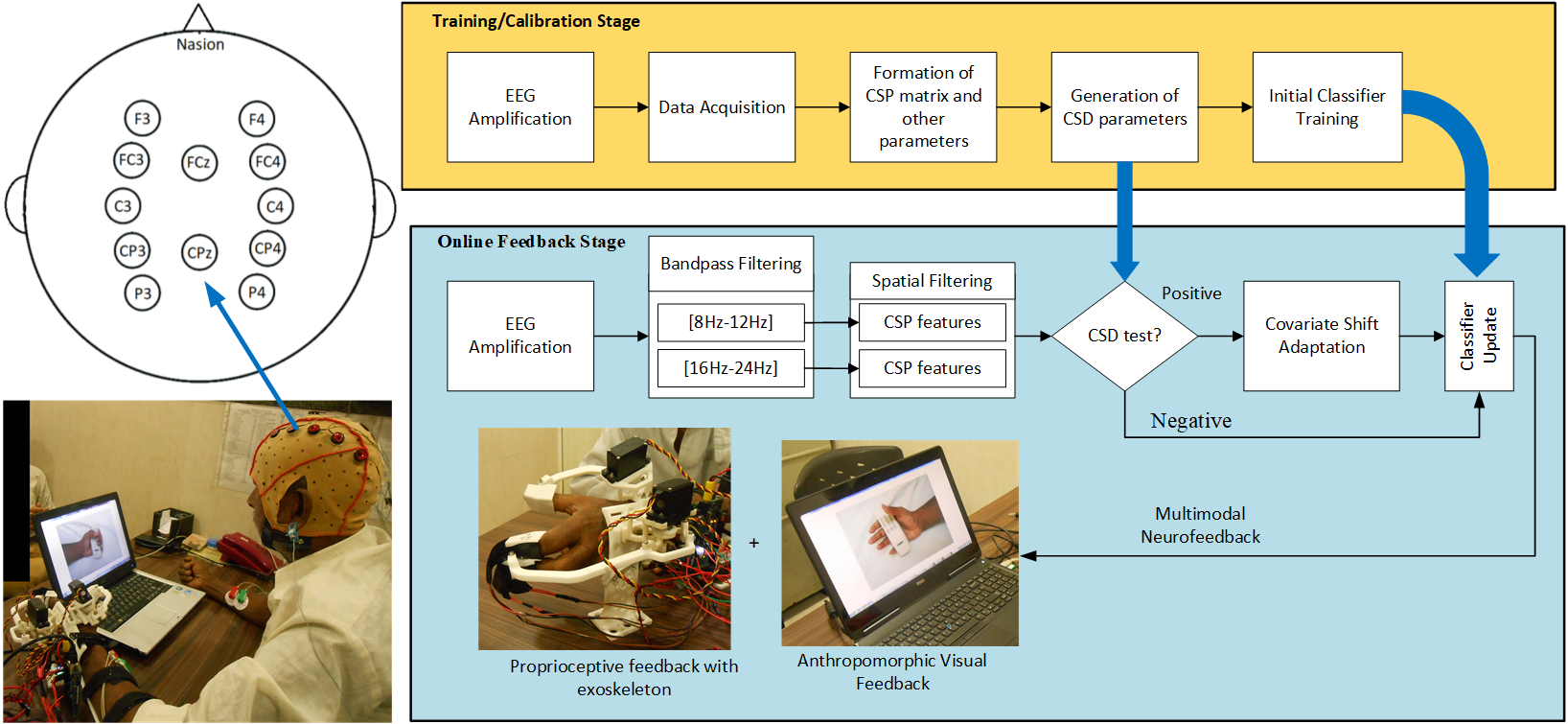}	
	\caption{The signal processing architecture is shown in the right half of the figure while the electrode placement is shown in the upper left corner and the experimental setup is shown in the lower left corner.}
	\label{fig:sigProArchitecture}
\end{figure*}


\begin{table*}[htb]
\caption{Demographics of the participants}
\centering
\label{tab:demo}
\begin{tabular}{|l|l|c|}
\hline
\multicolumn{1}{|c|}{\textbf{Institute}}                                 & \multicolumn{1}{c|}{\textbf{Country}} & \textbf{Number of Teams} \\ \hline
Institute of   Automation Chinese Academy of Sciences                    & China                                 & 1                    \\ \hline
Indian   Institute of Technology Kanpur                                  & India                                 & 2                    \\ \hline
Aramis   project-team, Inria Paris, Paris Brain Institute, Paris, France & France                                & 1                    \\ \hline
Federal   University of Juiz de Fora                                     & Brazil                                & 1                    \\ \hline
Huazhong   University of Science and Technology                          & China                                 & 1                    \\ \hline
Karadeniz   Technical University                                         & Turkey                                & 1                    \\ \hline
Qazvin Islamic   Azad University                                         & Iran                                  & 1                    \\ \hline
National   University of Science and Technology (NUST), Pakistan         & Pakistan                              & 1                    \\ \hline
IT Universe   Ltd., Samara, Russia                                       & Russia                                & 1                    \\ \hline
Technical   University of Kosice                                         & Slovakia                              & 1                    \\ \hline
The LNM   Institute of Information Technology                            & India                                 & 1                    \\ \hline
National   Institute of Technology, Warangal                             & India                                 & 1                    \\ \hline
\end{tabular}
\end{table*}
\begin{table*}[htb]
\centering
\caption{Results of the Overall category}
\label{tab:leagueTableOverall}
\begin{tabular}{|c|c|c|c|c|c|c|}
\hline
\textbf{}     & \textbf{}                                                    & \multicolumn{2}{c|}{\textbf{Within-Subject}} & \multicolumn{2}{c|}{\textbf{Within-Subject}} & \textbf{Overall} \\ \hline
\textbf{Rank} & \textbf{Institute}                                           & \textbf{Rank(R1)}    & \textbf{kappa($\kappa_1$)}    & \textbf{Rank(R2)}    & \textbf{kappa($\kappa_2$)}    & \textbf{os}      \\ \hline
\textbf{1}    & Federal University of   Juiz de For a, Brazil                & 7                    & 0.360                 & 1                    & 0.900                 & 15.480           \\ \hline
\textbf{2}    & IIT Kanpur, Kanpur,   India                                  & 3                    & 0.470                 & 2                    & 0.680                 & 14.480           \\ \hline
\textbf{3}    & Institute   of Automation Chinese Academy of Sciences, China & 5                    & 0.410                 & 4                    & 0.420                 & 8.720            \\ \hline
\end{tabular}
\end{table*}
\begin{table*}[htb]
\centering
\caption{Results of the cross-subject category}
\label{tab:leagueTableCross}
\begin{tabular}{|c|p{5cm}|c|c|p{5cm}|}
\hline
\textbf{Rank} & \multicolumn{1}{c|}{\textbf{Institute}}      & \textbf{accuracy (\%)} & \textbf{kappa} & \textbf{Method} \\ \hline
\textbf{1}    & Federal University of   Juiz de Fora, Brazil & 95.00                  & 0.90          & Filter-bank Common Spatial Pattern $+$ Single Electrode Energy features (SEE) $+$ Evolutionary optimization\\ \hline
\textbf{2}    & IIT Kanpur, Kanpur,   India                  & 83.75                  & 0.68           &
Riemmanian manifold (of signal covariance matrix) $+$ EEGNet \\ \hline
\textbf{3}    & Technical University   of Kosice, Slovakia   & 75.00                  & 0.50           &
1-D Resnet (9 layers) with squeeze and excitation blocks\\ \hline
\end{tabular}
\end{table*}
\begin{table*}[htb]
\centering
\caption{Results of the within-subject category}
\label{tab:leagueTableWithin}
\begin{tabular}{|c|p{5cm}|c|c|p{5cm}|}
\hline
\textbf{Rank} & \textbf{Institute}                                                       & \textbf{accuracy (\%)} & \textbf{kappa} & \textbf{Method} \\ \hline
\textbf{1}    & Aramis   project-team, Inria Paris, Paris Brain Institute, Paris, France & 78.44                  & 0.57           & Riemmanian manifold (of signal covariance matrix) $+$ functional connectivity features $+$ ensemble learning \\ \hline
\textbf{2}    & The   LNM Institute of Information Technology, Jaipur, India                     & 74.69                  & 0.49
& Common Spatial Patterns $+$ Support Vector Machines\\ \hline
\textbf{3}    & IIT Kanpur, Kanpur, India                                                        & 73.75                  & 0.47  & Riemmanian manifold (of signal covariance matrix) $+$ EEGNET        \\ \hline
\end{tabular}
\end{table*}
\begin{table}[htb]
\centering
\caption{Subject wise results for rank 1 to 3 in \textbf{within-subject} category}
\label{tab:subjectWiseWithinsubResults}
\begin{tabular}{|c|l|l|l|l|l|l|}
\hline
\multirow{2}{*}{\textbf{Subject-ID}} & \multicolumn{2}{c|}{\textbf{Rank1}} & \multicolumn{2}{c|}{\textbf{Rank2}} & \multicolumn{2}{c|}{\textbf{Rank3}} \\ \cline{2-7} 
                                 & \textbf{Acc(\%)}  & \textbf{$\kappa$}  & \textbf{Acc(\%)}  & \textbf{$\kappa$}  & \textbf{Acc(\%)}  & \textbf{$\kappa$}  \\ \hline
\textbf{P01}                     & 67.50             & 0.35            & 85.00             & 0.70            & 65.00             & 0.30            \\ \hline
\textbf{P02}                     & 92.50             & 0.85            & 80.00             & 0.60            & 82.50             & 0.65            \\ \hline
\textbf{P03}                     & 87.50             & 0.75            & 70.00             & 0.40            & 85.00             & 0.70            \\ \hline
\textbf{P04}                     & 77.50             & 0.55            & 67.50             & 0.35            & 67.50             & 0.35            \\ \hline
\textbf{P05}                     & 82.50             & 0.65            & 60.00             & 0.20            & 90.00             & 0.80            \\ \hline
\textbf{P06}                     & 50.00             & 0.00            & 70.00             & 0.40            & 50.00             & 0.00            \\ \hline
\textbf{P07}                     & 92.50             & 0.85            & 82.50             & 0.65            & 87.50             & 0.75            \\ \hline
\textbf{P08}                     & 77.50             & 0.55            & 82.50             & 0.65            & 62.50             & 0.25            \\ \hline
\textbf{Mean}                    & \textbf{78.44}             & \textbf{0.57}            & \textbf{74.69}             & \textbf{0.49}            & \textbf{73.75}             & \textbf{0.48}            \\ \hline
\textbf{Std}                     & 14.26             & 0.29            & 9.01              & 0.18            & 14.45             & 0.29            \\ \hline
\end{tabular}
\end{table}
\begin{table}[htb]
\centering
\caption{Subject wise results for rank 1 to 3 in \textbf{cross-subject} category}
\label{tab:subjectWiseCrossSubResults}
\begin{tabular}{|c|l|l|l|l|l|l|}
\hline
\multirow{2}{*}{\textbf{Subject-ID}} & \multicolumn{2}{c|}{\textbf{Rank1}} & \multicolumn{2}{c|}{\textbf{Rank2}} & \multicolumn{2}{c|}{\textbf{Rank3}} \\ \cline{2-7} 
                                 & \textbf{Acc(\%)}  & \textbf{$\kappa$}  & \textbf{Acc(\%)}  & \textbf{$\kappa$}  & \textbf{Acc(\%)}  & \textbf{$\kappa$}  \\ \hline
\textbf{P09}                     & 95.00             & 0.90            & 92.50             & 0.85            & 52.50             & 0.05            \\ \hline
\textbf{P10}                     & 95.00             & 0.90            & 75.00             & 0.50            & 97.50             & 0.95            \\ \hline
\end{tabular}
\end{table}

\begin{table*}[htb]
\caption{Accuracy comparison of impaired hand vs. healthy hand}
\centering
\label{tab:accCompImpvsHealthy}
\begin{tabular}{|c|c|c|}
\hline
\textbf{Institute}                                                       & \textbf{Impaired Hand Acc (\%)} & \textbf{Healthy Hand Acc (\%)} \\ \hline
Aramis project-team, Inria Paris, Paris Brain Institute, Paris,   France & 78.75                           & 78.125                         \\ \hline
The LNM Institute of Information Technology, Jaipur, India               & 70.625                          & 78.75                          \\ \hline
IIT Kanpur, Kanpur, India                                                & 73.125                          & 74.375                         \\ \hline
Federal   University of Juiz de For a, Brazil                            & 66.875                          & 69.375                         \\ \hline
Institute of Automation Chinese Academy of Sciences, China               & 86.25                           & 54.375                         \\ \hline
\end{tabular}
\end{table*}


\begin{table*}[htb]
\caption{Comparison of different features of the past competitions with CBCIC2020.}
\centering
\label{tab:compSOT}

\begin{tabular}{|c|c|c|c|c|c|c|c|c|c|c|}
\hline
\textbf{Competition} & \textbf{Dataset} & \textbf{Modality} & \textbf{Paradigm} & \textbf{\begin{tabular}[c]{@{}c@{}}\# of \\ Classes\end{tabular}} & \textbf{\begin{tabular}[c]{@{}c@{}}\# of \\ subjects\end{tabular}} & \textbf{\begin{tabular}[c]{@{}c@{}}\# training \\ trials\end{tabular}} & \textbf{\begin{tabular}[c]{@{}c@{}}\# of testing \\ trials\end{tabular}} & \textbf{\begin{tabular}[c]{@{}c@{}}\# of electrodes\\ or channels\end{tabular}} & \textbf{\begin{tabular}[c]{@{}c@{}}collection \\ mode\end{tabular}} & \textbf{\begin{tabular}[c]{@{}c@{}}\# of\\ Submissions\end{tabular}} \\ \hline
I                    & 1                & EEG               & SMR-ME            & 2                                                                 & 1                                                                  & 413                                                                    & 100                                                                      & 27                                                                              & open-loop                                                           & 6                                                                    \\ \hline
I                    & 2                & EEG               & SMR-MI            & 2                                                                 & 9                                                                  & 90                                                                     & 90                                                                       & 59                                                                              & open-loop                                                           & 3                                                                    \\ \hline
I                    & 3                & EEG               & SMR-MI            & 4                                                                 & 3                                                                  & 1152                                                                   & 768                                                                      & 64                                                                              & closed loop                                                         & 1                                                                    \\ \hline
II                   & Ia               & EEG               & SMR-MI            & 2                                                                 & 1                                                                  & 168                                                                    & 100                                                                      & 7                                                                               & closed loop                                                         & 15                                                                   \\ \hline
II                   & Ib               & EEG               & SMR-MI            & 2                                                                 & 1                                                                  & 200                                                                    & 200                                                                      & 7                                                                               & closed loop                                                         & 8                                                                    \\ \hline
II                   & IIa              & EEG               & SMR-MI            & 2                                                                 & 3                                                                  & 192                                                                    & 128                                                                      & 64                                                                              & closed loop                                                         & 5                                                                    \\ \hline
II                   & III              & EEG               & SMR-MI            & 2                                                                 & 1                                                                  & 280                                                                    & 280                                                                      & 3                                                                               & closed loop                                                         & 9                                                                    \\ \hline
II                   & IV               & EEG               & SMR-ME            & 2                                                                 & 1                                                                  & 316                                                                    & 100                                                                      & 28                                                                              & open-loop                                                           & 15                                                                   \\ \hline
III                  & I                & ECoG              & SMR-MI            & 2                                                                 & 1                                                                  & 278                                                                    & 100                                                                      & 64                                                                              & open-loop                                                           & 27                                                                   \\ \hline
III                  & IIIa             & EEG               & SMR-MI            & 4                                                                 & 3                                                                  & 240                                                                    & x                                                                        & 60                                                                              & unknown                                                             & 4                                                                    \\ \hline
III                  & IIIb             & EEG               & SMR-MI            & 2                                                                 & 3                                                                  & unknown                                                                & unknown                                                                  & 2                                                                               & unknown                                                             & 7                                                                    \\ \hline
III                  & IVa              & EEG               & SMR-MI            & 3                                                                 & 5                                                                  & 28-168                                                                 & 56-252                                                                   & 118                                                                             & unknown                                                             & 14                                                                   \\ \hline
III                  & IVb              & EEG               & SMR-MI            & 3                                                                 & 5                                                                  & unknown                                                                & unknown                                                                  & 118                                                                             & unknown                                                             & 1                                                                    \\ \hline
III                  & IVc              & EEG               & SMR-MI            & 3                                                                 & 5                                                                  & unknown                                                                & unknown                                                                  & 118                                                                             & unknown                                                             & 7                                                                    \\ \hline
III                  & V                & EEG               & Mix Imagery       & 3                                                                 & 3                                                                  & unknown                                                                & unknown                                                                  & 32                                                                              & open-loop                                                           & 26                                                                   \\ \hline
IV                   & 1                & EEG               & SMR-MI            & 3                                                                 & 4                                                                  & 200                                                                    & 240                                                                      & 59                                                                              & unknown                                                             & 24                                                                   \\ \hline
IV                   & 2a               & EEG               & SMR-MI            & 4                                                                 & 9                                                                  & 288                                                                    & 288                                                                      & 22                                                                              & open-loop                                                           & 5                                                                    \\ \hline
IV                   & 2b               & EEG               & SMR-MI            & 2                                                                 & 9                                                                  & 120                                                                    & 120                                                                      & 3                                                                               & closed loop                                                         & 6                                                                    \\ \hline
IV                   & 3                & MEG               & SMR-MI            & 4                                                                 & 2                                                                  & 160                                                                    & x                                                                        & 10                                                                              & unknown                                                             & 4                                                                    \\ \hline
IV                   & 4                & ECoG              & SMR-ME            & 5                                                                 & 3                                                                  & 150                                                                    & x                                                                        & 64                                                                              & unknown                                                             & 5                                                                    \\ \hline
\textbf{CBCIC2020}   & \textbf{1}       & \textbf{EEG}      & \textbf{SMR-MA}   & \textbf{2}                                                        & \textbf{10}                                                        & \textbf{80}                                                            & \textbf{40}                                                              & \textbf{12}                                                                     & \textbf{closed loop}                                                & \textbf{13}                                                          \\ \hline
\end{tabular}
\end{table*}




\subsection{Outcome}
A total of 14 teams participated in the competition from 12 Institutions across 9 different countries spread across 3 continents. The geographic distribution of the participants is shown in the map provided in Fig.~\ref{fig:map} and the demographics of the participants are shown in Table~\ref{tab:demo}. The winner of the within-subject category was Marie-Constance Corsi and colleagues
from Aramis project-team, Inria Paris, Paris Brain Institute with an average accuracy of 78.44\% and $\kappa$= 0.57. The second and third positions were achieved by Pramod Gaur and colleagues from The LNMIIT, Jaipur, India and Tharun Kumar Reddy
 and colleagues from IIT Kanpur, Kanpur, India, respectively. The accuracy in second position was 74.69\% ($\kappa$= 0.49) while the accuracy in third position was 73.75\% ($\kappa$= 0.47). In the cross-subject category, the winner was Gabriel Henrique de Souza and their team from Federal University of Juiz de Fora, Brazil, with 95\% accuracy ($\kappa$= 0.9). The second position was achieved by Tharun Kumar Reddy, and their team from IIT Kanpur with an accuracy 83.75\% ($\kappa$= 0.68) and the third position was achieved by Andrinandrasana David Rasamoelina and their team from Technical University of Kosice, Slovakia, with accuracy 75.00\% ($\kappa$= 0.50). There was also a third category which was called the "overall" category, which combined the performances from the first two categories (within-subject and cross-subject categories) and tried to estimate which submissions were better overall. The formula used to get this ranking is given below:
 
\begin{equation} 
 \label{eq:overallscore}
os = ((N+1)-R_w).\kappa_{w}+((N+1)-R_c).\kappa_{c}
\end{equation}

In Eq.(\ref{eq:overallscore}), $os$ represents the overall score, $\kappa_{w}$ and $\kappa_{c}$ are the kappa values of a particular submission in within-subject and cross-subject categories. The $R_w$ and $R_c$ are the corresponding rankings in within- and cross-subject categories, respectively, while $N$ is the total number of submissions. On the basis of $os$ the winner in the overall category was Henrique de Souza and colleagues, while the second and third positions were achieved by Tharun Kumar Reddy and colleagues and Long Cheng and colleague from the Institute of Automation Chinese Academy of Sciences, China. The details of the performance of the first, second, and third position holders in different categories are given in Table~,\ref{tab:leagueTableOverall} Table~\ref{tab:leagueTableCross} and Table~\ref{tab:leagueTableWithin} respectively. The comparison of performance between the within- and cross-subject categories is given in the bar graph of kappa values for the first 5 rank holders in Fig.~\ref{fig:interCatComp}.

\section{Dataset}
In this section, the dataset used in the competition is described along with the description of the subjects involved in the dataset, the experimental paradigm, and the data format. The link to download this open source dataset is as follows: https://github.com/5anirban9/Clinical-Brain-Computer-Interfaces-Challenge-WCCI-2020-Glasgow.git.

\subsection{Subjects}

The EEG data from 10 hemiparetic stroke patients were recorded. The average age across all the subjects was 47.5$\pm$15.31, while 6 of them were male and 4 female. Among the 10 subjects, 3 of them were impaired by their right hand side, and the rest of the 7 were impaired by their left hand. The average time since the first occurrence of stroke was 11$\pm$14.03 months. The subjects didn't have any prior experience of using a BCI system and had no record of epilepsy. They signed written informed consent forms before undergoing the experimentation. The experimental protocol was approved by the Institute Ethics Committee of the Indian Institute of Technology Kanpur.

\subsection{Experimental Protocol}
The data recording was divided into two sessions: first, the training session without feedback, and the next is the online feedback session. The training session consists of two runs of 40 trials each and the online feedback session consists of one run of 40 trials. Each trial started with a "get ready" message for 3~s, followed by a cue to perform the left or right hand grasp attempt, which lasted for 5~s during the training trials. There was a beep sound just 1~s before the appearance of the cue to alert the subjects. In the online feedback trials, the cue lasted for about 2~s and then the feedback was issued for the rest of the 3~s. Each trial both in training and in online feedback, lasted for 8~s, after which an inter-trial-interval (ITI) was there for about 2~s to 3~s. The timing diagram of a trial is given in Fig.~\ref{fig:timingDiagramTrial}. The trials are equally distributed into left and right within a run, i.e. 20 trials were for left hand task and 20 were for right hand task, which are generated randomly. Thus there were a total of 80 trials for training the classifier (generated during the 2 training runs) and 40 trials for testing (40 trials in the online feedback stage). The multimodal feedback issued during the online feedback stage was composed of visual feedback on a computer screen in the form of an anthropomorphic hand grasp and proprioceptive feedback mediated by a hand exoskeleton device. The hand-exoskeleton was 3-D printed with a nylon based material and powered by two servo motors to control the two arms of the exoskeleton responsible for moving the coupled index-middle fingers and the thumb. The exoskeleton weighs less than 600 g, and it was fully portable and wearable for a long time. The exoskeleton is capable of providing assistance to stroke patients to do the flexion and extension motion of their fingers. The servo motors are coupled with a four-bar mechanism that drives the joints in such a way that the fingertips follow the natural human finger trajectory so that any unnecessary pressure on the fingers can be avoided~\cite{Chowdhury2019IEEETH}. The experimental setup is depicted in Fig~\ref{fig:sigProArchitecture} in the bottom left corner.

\subsection{Data Acquisition}
The EEG data were recorded from 12 electrodes, mostly over the motor cortex, which are: F3, FC3, C3, CP3, P3, FCz, CPz, P4, FC4, C4, CP4, and P4 according to 10-20 international system. The placement of the electrodes is shown in Fig.~\ref{fig:sigProArchitecture} in the top left corner. The data acquisition hardware was g.USBamp (g.tec, Graz, Austria) biosignal amplifier, Ag/AgCl-based EEG electrodes, and EEG cap (g.tec). The EEG signals were recorded at 512~Hz sampling rate and initially bandpass filtered between 0.1~Hz and 100~Hz with a notch filter at 50~Hz for cancelling the power line noise. The software for data acquisition, signal processing and feedback generation was built in-house in a MATLAB/Simulink$^{TM}$. The MATLAB based software was connected to the Arduino based controller of the exoskeleton to synchronize its movement with the visual feedback.

\subsection{Data format}
The dataset was shared through a GitHub repository. There are two files for each subject. The file name ending with `T' designates the training file, and the file name ending with `E' designates the evaluation/testing file. For example, the filename `Parsed\_P05T' suggests the training file for subject P05, while `Parsed\_P05E' suggest the evaluation/testing file for the same subject. The training files contains the labels corresponding to each trial, while the labels for trials of evaluation/testing files are not provided. The objective of the competition is to find the labels corresponding to the trials of the evaluation/testing files. All the files are in `.mat' (MATLAB) format, so they can easily be opened using MATLAB software. A training file for any subject (for example, file `Parsed\_P05T', which is the training file for subject `P05') contains two variables `rawdata' and `labels'. The variable `rawdata' is a 3-D matrix of dimension $T\times C\times S$ where $T$ denotes the number of trials, $C$ denotes the number of channels, and $S$ denotes the number of samples. Here, in all the files $T$ = 80, which means there are a total of 80 trials in the training file. Next, in all the files $C$=12, which means there were 12 EEG channels during the recording. The index of the channels from 1 to 12 denotes the channels F3, FC3, C3, CP3, P3, FCz, CPz, P4, FC4, C4, CP4, and P4 in the same order. This means index1= F3, index2= FC3, and so on up to index12= P4. Next, the number of samples per trial $S$= 4096. The explanation for this is that each trial is 8 s long, and the data was recorded with a sampling rate of 512 Hz. So, 8$\times$512 = 4096. Thus rawdata(5, 10, :) contains the information about the activity of EEG channel C4 at 5$-$th trial, rawdata(10, 5, :) contains the information about the activity of EEG channel P3 at 10$-$th trial and so on. Now, the variable `labels' is a 1-D array of dimension 80$\times$1 containing the labels for individual trials in the training data. For example, label(1) contains the label for trial index 1 and label(80) contains the label for trial index 80. As there are two classes, namely `left motor attempt' and `right motor attempt', labels are either '1' or '2', where '1' corresponds to the `right motor attempt', and 2 corresponds to the `left motor attempt'. The same explanation goes for the evaluation or testing files with the exceptions that the $T$=40, which means the dimension of raw data will be 40$\times$12$\times$4096, and there would be no `labels' as it is to be predicted by the classifier trained on the training data.

\subsection{Signal Processing}
The EEG data and class labels acquired during the training phase were used to extract the features based on common spatial patterns (CSP) to train a support vector machine based (SVM) classifier using a linear kernel in a supervised learning architecture. Additionally, covariate shift detection (CSD) parameters were also calculated from the training data, which would later be used to detect the nonstationary changes in the data distribution and adapt the classifier accordingly. In the online feedback stage, the prediction of the class labels for a particular trial starts with the same classifier trained after the training phase, but it was then updated upon the detection of each valid covariate shifts as the trials progress. Once a covariate shift is detected, the newly available information from the online feedback phase is appended with the initial training data, and the classifier is adapted accordingly to determine new decision boundaries. The flow diagram of the signal processing architecture is shown in Fig.~\ref{fig:sigProArchitecture}. The details of the signal processing steps can be found in~\cite{Chowdhury2018IEEETCDS}. 

\subsection{Winner Methods}

The competition participants are also requested to share their methods and algorithms along with the predicted labels. Some of these methods and algorithms are described here. The submissions cover various methods ranging from traditional methods such as filter bank CSP (FBCSP) to emerging methods such as Riemannian geometry and Convolutional Neural Networks (CNN). We will explain further the results on the winner methods for each category.

\subsubsection{Best within-subject category}
The method which achieved the best accuracy in the within-subject category was the Riemannian geometry based classifier. However, the authors Marie-Constance Corsi and colleagues used the functional connectivity based features in addition to the classical covariance matrix. They also used ensemble learning along it for robust classification. Here, they have explored the association of complementary undirected functional connectivity measures with Riemannian geometry. The frequency window for extracting the features was 8-30~Hz, while the corresponding time window was 3 to 7.5~s within the trial. For the final prediction, the ridge classifier was used, which took input from an ensemble of FgMDM classifier~\cite{Barachant2010Springer} used for classifying individual features. The inter-session shift in the data distribution between training and testing is handled by transporting the test data to the training data mean as described in~\cite{Barachant2012}.


\subsubsection{Best cross-subject category}
The winning submission for the cross-subject category from Gabriel Henrique de Souza and colleagues used the FBCSP and Single Electrode Energy (SEE) with best parameter selection to predict class labels. They found FBCSP as limited in capacity for the prediction in the cross-subject case and hence, developed a new method called SEE, which can be used in both within- and cross-subject cases. The SEE method uses the signal from only one electrode, which is then passed through a temporal filter and a function to calculate the characteristic feature value. The trials are then ordered by these characteristic feature values, which are then split by the median value in order to assign the two different classes to the two different halves as it was a balanced classification problem. For FBCSP implementation, a 2~s time window within the trial was used for feature extraction while 9 temporal filters were there. The optimal features were selected using the MIBIF method~\cite{Keng2012}. However, ultimately SEE was used for all the subjects except P06. The suitable SEE passband frequency for P09, and P10 was described as 20-36~Hz, while the single electrode used for classification was CPz.

\subsubsection{Best balanced inter- and cross-subject method}
An interesting observation from the results is that the methods that performed well in within-subject case do not necessarily mean that they performed well in cross-subject case. But there were two submissions that performed reasonably well in both categories. The first one is the submission from Tharun Kumar Reddy and colleagues who achieved 3rd position in within subject category and 2nd position in cross-subject category. They experimented mainly with two approaches, namely End-to-end classification of raw EEG signals using Convolutional Neural Network (CNN) and Classification using Riemannian space of covariance matrices. The CNN architecture used in this case is the EEGNet architecture proposed by Lawhern and colleagues in 2016~\cite{Lawhern2018}. For the classification in within-subject category they used the Riemannian geometry (Rg) features with Dense Neural Network (DNN) classifier. The model Rg+DNN model was chosen based on the 5-fold cross validation on the training data. In the case of cross-subject decoding, an ensemble of Rg+DNN, Rg+SVM, and Rg+EEGNet was used while the final prediction was done using the technique of majority voting.

Another method that yielded more than 70\% accuracy both in the case of within- and cross-subject categories while achieved 3rd position in the overall category was the submission from Long Cheng and colleagues. They developed a multimodel convolutional neural network (MCNN) having 4 sub-models with different model structures. The input to the model was bandpass filtered between 7Hz and 36 Hz, while sliding window based data augmentation was used to overcome the effects of a limited number of trials. New convolutional kernels such as DepthwiseConv2D and SeparableConv2D are also developed for the efficient performance of the MCNN architecture.

In order to see if various algorithms have different recognition rates between the impaired hand and the healthy hand, we calculated the accuracies for both hands separately and grouped them into impaired hand and healthy for all the top ranking submissions. This can be seen in Table~\ref{tab:accCompImpvsHealthy} which shows the effect of impairment on the classification. 

\section{Discussion}
The success of the Clinical BCI Challenge (CBCIC2020) can be evaluated in many different aspects. The most important one is perhaps making a good quality stroke patients' dataset  openly available which was recorded in a neurorehabilitative BCI paradigm. A major application area in BCI is using such interfaces to control robotic devices purposed for doing physical exercises as a part of rehabilitation therapy~\cite{RatheeChowdhury2019IEEETNSRE,Chowdhury2019JNM}. Therefore, when a new algorithm is proposed, the researchers are often interested in evaluating its efficiency to decode sensorimotor rhythms associated with motor intentions. However, such studies are often considered incomplete unless they are tested on stroke patients whose neurodynamics is generally altered as compared to the healthy individuals due to the presence of a lesion. The inter-session and inter-subject non-stationarity are also greater in the case of the subjects with stroke with an additional adverse effect of low SNR. As the datasets shared in the past BCI Competitions mostly dealt with healthy individuals newly proposed algorithms are often untested on the actual end-users of BCI who are largely in the clinical domain. Recording the data from stroke patients is also a very challenging task as it involves time consuming ethical approval processes, especially when medical devices such as exoskeletons are involved, and due to the difficult patient recruitment process. Thus, CBCIC2020 intended to fill this void so that a systematic benchmarking of the past and emerging methodologies can be possible for their potential use in neurorehabilitation. 

Unlike some of the past datasets where a number of subjects are not sufficient to determine statistical significance (such as BCI Competition II-Ia dataset with one ALS patient), the CBCIC2020 consists of 10 stroke patients, sufficient to evaluate the statistical significance while comparing two different algorithms. The number of subjects ($n=$10) is also comparable with other popular EEG based MI-BCI datasets such as BCI Competition IV-2a and 2b, which both have 9 subjects. Hence, an algorithm that is already evaluated in such healthy individuals' dataset can be compared further for its performance on stroke patients using CBCIC2020 dataset. Another advantage of the CBCIC2020 dataset is that it is recorded in a closed-BCI paradigm similar to most previous competition datasets. This means that the methodologies evaluated in this dataset could be potentially used for online BCI feedback, which is an important criterion for its practical feasibility. 

The number of submissions received for this competition (total submission = 13) is also greater than the average number of submissions per dataset (average submission = 9.5) in BCI Competion I-IV. The submissions cover a large variety of algorithms with sufficiently high performance, which also indirectly verifies the signal quality of the dataset. Interestingly, it can be seen that the signal processing pipelines, methods, or algorithms that work best in one category (within-subject/cross-subject) don't necessarily mean that they are also the best for the other. If we look at submissions from the top performers, we can see that the SEE method, which worked very well for the cross subject prediction didn't work well for within-subject. The Riemannian geometry based approach with functional connectivity and ensemble learning was the best performer for within-subject category but not in the case of cross-subject. On the other hand the same Riemannian geometry based approach with a combination of different neural network architectures (such as DNN and EEGNet) performed almost equally well in within- and cross-subject categories. Other methods such as the traditional CSP+SVM based approach seemed to be more suited for the within-subject classification, while the results in cross-subject classification were not good enough. The newly proposed MCNN approach gave good accuracy (more than 70\%) in both the categories, which could be advantageous as it focuses on end-to-end learning. As compared to the previous within-subject results on the same dataset using different algorithms (such as CSP+SVM~\cite{Chowdhury2018IEEETCDS} and EEGNet~\cite{RazaChowdhuryEEGNet2020IJCNN}) the winner of the competition performed better. For example, the average accuracy across P01 to P08 in~\cite{Chowdhury2018IEEETCDS} and~\cite{RazaChowdhuryEEGNet2020IJCNN} 70$\pm$3.78\% and 70$\pm$15.86\% respectively, while the same for the winner of the competition was 78.44$\pm$14.26\%. On the downside, it is also important to note that the inter-subject variability in performance for P01 to P08 is high in all the top 3 submissions as the standard deviation 14.26\%, 9.01\%, and 14.45\% respectively for rank1, rank2, and rank3. This shows that the traditional methods such as CSP+SVM although a bit low in average accuracy has lower inter-subject variability in performance as revealed by the standard deviation. It is important to reduce the standard deviation to get a statistically significant difference in performance, which can be a target for future algorithms. If we look at the cross-subject results, we can see that the best accuracy achieved for P09 and P10 were both at 95\%, which is an unprecedented performance considering no training data were given for those two subjects. Previously, the accuracies achieved for P09 and P10 were 72.5\% and 70\% respectively using CSP+SVM~\cite{Chowdhury2018IEEETCDS}, and 52.5\% and 90\% respectively using EEGNet~\cite{RazaChowdhuryEEGNet2020IJCNN}, although they used the training data of P09 and P10. The cross-subject performance achieved by 2nd rank holder was also greater than~\cite{Chowdhury2018IEEETCDS}, only the P10 accuracy was greater in~\cite{RazaChowdhuryEEGNet2020IJCNN}. The P09 accuracy for 3rd rank holder was comparable to~\cite{RazaChowdhuryEEGNet2020IJCNN} at chance level only, while the P10 accuracy was greater at 97.5\%. In spite of the fact that the top rankers of the competition achieved better performance in the case of cross-subject decoding, it is to be noted that the robustness of the algorithms can only be evaluated in a leave-one-out setting where the accuracies and kappa values would be calculated for every subject (P01 to P10) from the training data of the rest of the subjects leaving the training data of that particular subject.

There are however some shortcomings of the dataset as compared to the previous ones. First of all, the number of electrodes used for recording the EEG data is relatively smaller (12 electrodes) while a large number of previous datasets (in Table~\ref{tab:compSOT}) used 32 to 64  electrodes. Thus there is limited scope of doing a rich analysis of brain-connectivity, which may need broad spatial coverage. The number of trials both in training (number of trials=80) and testing (number of trials=40) sessions are also much less than many previous SMR datasets  given in Table~\ref{tab:compSOT}, although learning from small examples is one of the key challenges in practical BCI. Another drawback is that the dataset only deals with 2 classes, and hence, an algorithm cannot be validated for its performance in multiclass BCI using this dataset.

\section{Conclusion}
For the first time, a BCI Competition combines two important aspects of SMR-BCI decoding: the within-subject and cross-subject classification involving stroke patients. It is particularly important at this hour as the major thrust in BCI research is to develop algorithms and techniques for calibration free BCI designs, which is essential for real-world applications. The quality participation from different corners of the world achieved sufficiently high performance in a challenging dataset involving stroke patients, which is very rare. Overall, CBCIC2020 gives a rich open dataset conforming with the standards of a neurorehabilitative BCI paradigm and provides required benchmarks encompassing various past and emerging algorithms that can act as a test bench for developing robust and efficient techniques for BCI based robotic rehabilitation. Ideas for future competitions may be about performing BCI in natural settings controlling remote devices \cite{cortez2021}, robots \cite{andreu2016} or in virtual reality \cite{achanccaray2017}.

\section*{Acknowledgment}
We are very thankful to the IEEE WCCI 2020 Competition Chairs and the IEEE Computational Intelligence Society for accepting and sponsoring this competion, aslo we want to thanks very much to Guger Technologies (g.tec medical engineering) for supporting the awards. The data was recorded as a part of a  DST-UKIERI project led by Prof. Girijesh Prasad (Ulster University, UK) and Prof. Ashish Dutta (IIT Kanpur, India), named ``A BCI Operated Hand Exoskeleton Based Neurorehabilitation System" under Grant UKIERI-DST-2013-14/126, Grant DST/INT/UK/P-80/2014 and Grant DST-UKIERI-2016-17-0128.

\ifCLASSOPTIONcaptionsoff
  \newpage
\fi



\bibliographystyle{IEEEtran}

\bibliography{refRevCBCIC2020}
%

%








\end{document}